\documentclass[sigconf]{acmart}

\usepackage[hypcap=false]{caption}
\usepackage{tikz}
\usepackage[framemethod=tikz]{mdframed}
\usepackage[normalem]{ulem}

\usetikzlibrary{arrows,automata,positioning}
\tikzset{inner sep=0pt,minimum size=1pt}

\newcommand{\now}{\mathit{now}}
\renewcommand{\epsilon}{\varepsilon}

\newtheorem{theorem}{Theorem}
\theoremstyle{definition}
\newtheorem{definition}{Definition}
\surroundwithmdframed[outerlinewidth=0.4pt,
  innerlinewidth=0.4pt,
  middlelinewidth=0pt,
  middlelinecolor=white]{definition}

\copyrightyear{2020}
\acmYear{2020}
\setcopyright{rightsretained}
\acmConference[SPAA '20] {32nd ACM Symposium on Parallelism in Algorithms and Architectures}{July 15--17, 2020}{Virtual Event, USA}
\acmBooktitle{Proceedings of the 32nd ACM Symposium on Parallelism in Algorithms and Architectures
  (SPAA '20), July 15--17, 2020, Virtual Event, USA}
\acmDOI{10.1145/3350755.3400264}
\acmISBN{978-1-4503-6935-0/20/07}
\acmSubmissionID{spaa91ba}
\begin{document}

\title[Feasibility of Cross-Chain Payment with Success Guarantees]
{{\LARGE Brief Announcement:}\texorpdfstring{\\}{}Feasibility of Cross-Chain Payment with Success Guarantees}
\author{Rob van Glabbeek}
  \affiliation{%
   \institution{Data61, CSIRO and\\ UNSW}
%  \city{Sydney}
%  \country{Australia}
}
\email{rvg@unsw.edu.au}
\author{Vincent Gramoli}
  \affiliation{%
   \institution{University of Sydney and\\ EPFL}
%  \city{Lausanne}
%  \country{Switzerland}
}
\email{vincent.gramoli@sydney.edu.au}
\author{Pierre Tholoniat}
  \affiliation{%
   \institution{University of Sydney and\\ \'Ecole Polytechnique}
%  \city{Paris}
%  \country{France}
}
\email{pierre.tholoniat@polytechnique.edu}

\begin{abstract}
  We consider the problem of cross-chain payment whereby customers of different
  escrows---implemented by a bank or a blockchain smart contract---successfully
  transfer digital assets without trusting each other. Prior to this work,
  cross-chain payment problems did not require this success, or any form of progress.
  We demonstrate that it is possible to solve this problem when assuming synchrony,
  in the sense that each message is guaranteed to arrive within a known amount of
  time, but impossible to solve without assuming synchrony. Yet, we solve a weaker
  variant of this problem, where success is conditional on the patience of the
  participants, without assuming synchrony, and in the presence of Byzantine failures.
  We also discuss the relation with the recently defined cross-chain deals.
\end{abstract}

\maketitle

\section{Introduction and Related Work}
        
With the advent of various payment protocols comes the problem of interoperability between them. 
A simple way for users of different protocols to interact is to do a \emph{cross-chain payment}
whereby intermediaries can help customer Alice transfer digital assets to Bob even though
Alice and Bob own accounts in different banks or blockchains.  
To implement a payment between customers of different banks, it helps if the two
banks have ways to transfer assets to each other, and moreover trust each other. 
The problem becomes more interesting when this is not the case.
Thomas and Schwartz~\cite{TS15} propose two cross-chain payment protocols:
(i)~the universal protocol requires \emph{synchrony}~\cite{DLS88};
(ii)~the atomic protocol merely requires \emph{partial synchrony}~\cite{DLS88}.
Herlihy, Liskov and Shrira~\cite{Her19bis} represent a cross-chain payment as a deal matrix $M$
where $M_{i,j}$ characterises a transfer of some asset from participant $i$ to participant $j$. 
They offer a timelock protocol that requires synchrony, and a certified blockchain protocol that requires partial synchrony.
However, the synchronous solutions of \cite{TS15} and \cite{Her19bis} do not consider clock
drift, and for their partially synchronous solutions no success guarantees are established.

In this brief announcement, we formally define the \emph{time-bounded cross-chain payment problem},
and show that, assuming synchrony, there exists an algorithm that solves it.  Our solution is the
universal protocol of \cite{TS15}, but fine-tuned to work correctly in the presence of clock drift.
We also prove that this problem cannot be solved when merely assuming partial synchrony, even if we
relax the problem statement by merely requiring eventual (instead of time-bounded) termination.
Moreover, inspired by earlier work on the transaction commit problem~\cite{Had90}, we define a
weaker variant of our problem that relaxes the liveness guarantee to be solvable with partial
synchrony.  Contrary to the problem statements in \cite{TS15} and \cite{Her19bis}, a protocol where
all participants always abort is not permitted by our problem specification.
\vspace{-0.5em}

\section{Model and definitions}\label{model}

We assume $n$ banks or \textit{escrows} $e_0,\dots,e_{n-1}$ and $n{+}1$ \textit{customers}
$c_0,\dots,c_{n}$.  These $2n{+}1$ processes are called \textit{participants}.  An escrow is a
specific type of process that can handle values for other parties in a predefined manner.  Customer
$c_0$ is Alice and $c_n$ is Bob. The customers $c_1,\dots,c_{n-1}$ are intermediaries in the
interaction between Alice and Bob; we call them \textit{connectors}, named\ Chloe$_i$.  Customers
$c_{i-1}$ and $c_{i}$ have accounts at escrow $e_{i-1}$, and trust this escrow ($i=1,\dots,n$).  We
do not assume any other relations of trust.
\vspace{1ex}

\noindent\begin{minipage}{\linewidth}
  \begin{center}
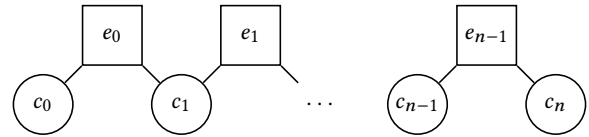

  \begin{tikzpicture}[xscale=0.45,yscale=0.75,>=latex', auto, semithick, node distance=1.3cm]
    
    \node[state, circle]   (A)                        {$c_0$};
    \node[state, rectangle]           (B) [above right of = A]        {$e_0$};
    \node[state, circle]           (C) [below right of = B]        {$c_1$};
    \node[state, rectangle]           (D) [above right of = C]        {$e_1$};
    
    \node[state, draw=white]           (E) [below right of = D]        {$\ldots$};
    
      \node[state, circle]   (F)          [right of = E]              {$c_{n-1}$};
    \node[state, rectangle]           (G) [above right of = F]        {$e_{n-1}$};
    \node[state, circle]           (H) [below right of = G]        {$c_n$};
  
    \path (A) edge                node        {} (B)
          (B) edge                node        {} (C)
          (C) edge                node        {} (D)
          (D) edge                node        {} (E)
          (F) edge                node        {} (G)
          (G) edge                node        {} (H)   
          ;
   
  \end{tikzpicture}
  \end{center}
  \captionof{figure}{Customers and escrows.}
  \label{fig:topology}
  \vspace{3ex}
  \end{minipage}

We assume that value can be transferred directly only between customers of the same escrow.
Moreover, any transfer between two customers of an escrow can be modelled as two transfers:
one from the originating customer to the escrow, and one from the escrow to the receiving customer.
Thus, the connections from Figure~\ref{fig:topology} describe both the relations of trust and the
possible transfers.

Two customers may make a deal with an escrow to place value from the first customer ``in escrow'',
and, after a predefined period, depending on which conditions are met, either complete the transfer
to the second customer, or return the value to the first one.

We suppose that the participants have already agreed upon the value they expect to transfer.
As Chloe helps out transferring value from Alice to Bob, it is only reasonable that she is paid a
small commission. Hence the value transferred from Alice to Chloe might be larger than the value
transferred from Chloe to Bob. Additionally, these values may be expressed in different currencies,
or they may be objects.  Deciding which values to transfer may thus be an interesting problem.
However, it is entirely orthogonal to the matter discussed here, and hence we shall not consider it any further.

We consider the classic Byzantine model with authentication.

\section{Feasibility of Cross-Chain Payments}

A cross-chain payment protocol prescribes a behaviour for each of the participants in the protocol, the
escrows and the customers. 
Let $\chi$ be a certificate signed by Bob saying that Alice's obligation to pay him has been met.
\vspace{3pt}

\begin{definition}
\vspace{-1ex}
  A protocol is a \textit{time-bounded cross-chain payment protocol} (resp. an \textit{eventually
  terminating cross-chain payment protocol}) if it satisfies the following properties:
  \setlength\leftmargini{24pt}
  \begin{itemize}\leftmargin 0pt
    \item[C] \textbf{Consistency.} For each participant in the protocol it is possible to abide by the protocol.
    \item[T] \textbf{Time-bounded (resp. eventual) termination.} \hfill Each customer that abides by the
      protocol, and either makes a payment or issues a certificate, terminates within an
      a priori known period (resp. terminates eventually), provided her escrows abide by the protocol.
    \item[ES] \textbf{Escrow security.} Each escrow that abides by the protocol does not lose money.
    \item[CS] \textbf{Customer security.}
      \begin{itemize}
        \item[CS1] Upon termination, if Alice and her escrow abide by the protocol, Alice  has
          either got her money back or received the certificate $\chi$.
        \item[CS2] Upon termination, if Bob and his escrow abide by the protocol, Bob has either
          received the money or not issued certificate $\chi$.
        \item[CS3] Upon termination, each connector that abides by the protocol has got her money back,
          provided her escrows abide by the protocol.
      \end{itemize}
    \item[L] \textbf{Strong liveness.} If all parties abide by the protocol, Bob is paid eventually.
  \end{itemize}
\end{definition}

Requirement C (\emph{consistency} of the protocol) is essential.
In the absence of this requirement, any protocol that prescribes an impossible task for each
participant would be a correct cross-chain payment protocol (since it trivially meets T, ES, CS and L).

Requirements ES and CS (the \emph{safety} properties) say that if a participant abides by the
protocol, nothing really bad can happen to her. These requirements do not assume that any other
participant abides by the protocol, and should hold no matter how malicious the other participants
turn out to be. The only exception to that is that the safety properties for a customer (CS)
are guaranteed only when the escrow(s) of this customer abide by the protocol.

Property L, saying that the protocol serves its intended purpose, is the only one that is
contingent on \emph{all} parties abiding by the protocol.
The proofs of the following results are presented in \cite{GGT19}.

\begin{theorem}\label{thm:possibility-sync}
If communications and computations are synchronous, there exists a time-bounded cross-chain payment protocol.
\end{theorem}

\begin{theorem}\label{thm:impossibility}
If communications are partially synchronous, there is no eventually terminating cross-chain payment protocol.
\end{theorem}

In view of this impossibility result we propose to weaken the liveness guarantee as
indicated in Def.~\ref{etccppwwlg}. In \cite{GGT19} we present a protocol in which each customer can, at
any moment of their choice, lose patience and abort the transaction, without a risk of losing
value. In case none of them exercises this option nor fails, a successful outcome is guaranteed.
This solution involves an external \emph{transaction manager}, that can issue an \emph{abort} or
\emph{commit} certificate. Properties CC and CS2 together guarantee that the commit
certificate can be used by Alice as a proof that Bob has been paid.
The transaction manager could be a single external party trusted by all,
or a smart contract running on a permissionless blockchain shared by every customer.
It can also be a collection of notaries appointed by the participants in the protocol, of which
less than one-third is assumed to be unreliable. They would run a consensus algorithm for
partial synchrony such as the one from Dwork, Lynch \& Stockmeyer \cite{DLS88}.%
\vspace{3pt}

\begin{definition}
\label{etccppwwlg}
\vspace{-1ex}
A protocol is a \textit{cross-chain payment protocol with weak liveness guarantees} if it satisfies
properties C, ES and CS3, as well as:
\begin{itemize}
\item[CC] \textbf{Certificate consistency.} An abort and a commit certificate can never be issued both.
\item[T] \textbf{Termination.} Each customer that abides by the protocol
  terminates eventually, provided her escrows abide by the protocol.
\item[CS] \textbf{Customer security.}
\begin{itemize}
\item[CS1] Upon termination, if Alice and her escrow abide by the protocol, Alice  has either got
  her money back or received the \textit{commit certificate} $\chi_c$.
\item[CS2] Upon termination, if Bob and his escrow abide by the protocol, Bob has either received
  the money \textit{or the abort certificate} $\chi_a$.
\end{itemize}
\item[L] \textbf{Weak liveness.} If all parties abide by the protocol and if the customers wait
  sufficiently long before and after sending money, then Bob is eventually paid.
\end{itemize}
\end{definition}

\begin{theorem}\label{thm:possibility}
There exists a cross-chain payment protocol with weak liveness guarantees.
\end{theorem}

\section{A time-bounded protocol}

\begin{figure}[t]
  \footnotesize
  \input{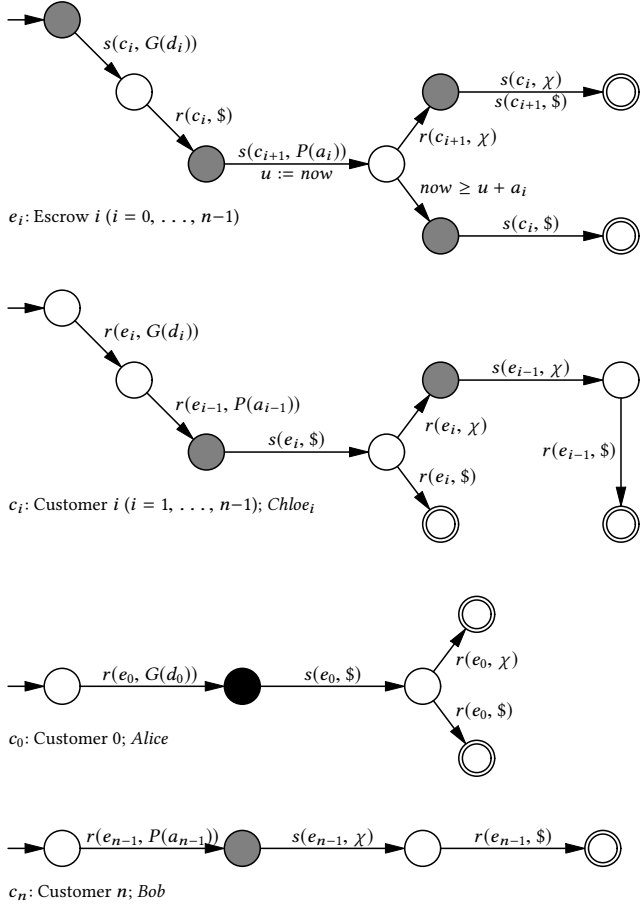}
  \centerline{\box\graph}
  \caption{Automata representing escrows and customers}
  \label{automata}
  \vspace{-4ex}
  \mbox{}
\end{figure}

In Figure~\ref{automata} we present the protocol from Thm.~\ref{thm:possibility-sync} formalised as
an Asynchronous Network of Timed Automata (ANTA), a specification formalism introduced in \cite{GGT19}.
There is one automaton for each participant in the protocol, that is, for each escrow $e_i$
($i=0,\dots,n{-}1$) and each customer $c_i$ ($i=0,\dots,n$). Each automaton is equipped with a
unique identifier, in this case $e_i$ and $c_i$.
It has a finite number of \emph{states}, depicted as circles, and \emph{transitions} between them.

Each automaton keeps an internal clock, whose value, a real number, is stored in the variable $\now$.
In case a transition occurs that is labelled by an assignment $x:=\now$,
the variable $x$ will remember the point in time when the transition took place.

An automaton spends a bounded amount of time calculating in each grey (\emph{output}) state, and leaves it by
performing the action $s(id,m)$ of sending message $m$ to participant $id$.
We consider three kinds of messages: (i) certificate $\chi$, signed by Bob,
(ii) the value $\$$ that is transmitted from one participant to another,
and (iii) promises made by escrow $e_i$ to its customers $c_i$ and $c_{i+1}$, respectively:%
\vspace{1ex}

\noindent
$G(d) :=$ ``\parbox[t]{2.9in}{I guarantee that if I receive $\$$ from you at my local time $w$,\\
              then I will send you either $\$$ or $\chi$ by my local time $w+d$.''}\vspace{1ex}

\noindent
$P(a) :=$ ``\parbox[t]{2.9in}{I promise that if I receive $\chi$ from you at my time $v$, with $v<\now+a$,
              then I will send you $\$$ by my local time $v+\epsilon$.''}\vspace{6pt}

When an automaton is in a white (\emph{input}) state, it stays there (possibly forever)
until one of its outgoing transitions becomes \emph{enabled}; in that case that transition will be taken immediately.
The \emph{time-out} transition $\now \geq u+a_i$ is enabled when this formula
evaluates to {\tt true}. An \emph{input} transition $r(id,m)$ is triggered
by the receipt of message $m$ from the automaton $id$ in the network.

The automata of Figure~\ref{automata} can be informally described as follows:
An escrow $e_i$ first sends promise $G(d_i)$ to its (upstream) customer $c_i$.
Here ``upstream'' refers to the flow of money.
The precise values of $d_i$ are calculated in \cite{GGT19}; here they are simply parameters in
the design of the protocol. Then it awaits receipt of the money/value from customer $c_i$.
If the money does arrive, the escrow issues promise $P(a_i)$ to its downstream customer $c_{i+1}$.
It remembers the time this promise was issued as $u$.
Then it awaits receipt of the certificate $\chi$ from customer $c_{i+1}$.
If $\chi$ does not arrive by time $u+a_i$, a time-out occurs, and the escrow refunds the money to
customer $c_i$. If it does arrive in time, the escrow reacts by forwarding the certificate to
customer $c_i$, and forwarding the money to customer $c_{i+1}$.

A connector Chloe$_i$ starts by awaiting promises $G(d_i)$ from her downstream escrow $e_i$, and
$P(a_{i-1})$ from her upstream escrow $e_{i-1}$. Then she proceeds by sending the money to
escrow $e_i$. After sending the money, Chloe$_i$ waits for escrow $e_i$ to send her either the
certificate $\chi$ or the money back. In the latter case, her work is done; in the former, she
forwards the certificate to escrow $e_{i-1}$ and awaits for the money to be sent by escrow $e_{i-1}$.
The automata for Alice and Bob are both simplifications of the one for Chloe$_i$.

\section{Relation with cross-chain deals}\label{deals}

In Herlihy, Liskov and Shrira~\cite{Her19bis}, a \emph{cross-chain deal} is given by a
matrix $M$ where $M_{i,j}$ is listing an asset to be transferred from party $i$ to party $j$.
It can also be represented as a directed graph, where each vertex represents a
party, and each arc a transfer; there is an arc from $i$ to $j$ labelled $v$ iff $M_{i,j} = v \neq 0$.
They present two protocols for implementing such a deal, while aiming to ensure:
\begin{itemize}
\item \textbf{Safety.}  For every protocol execution, every compliant party ends up with an acceptable payoff.
\item \textbf{Termination.}\footnote{In \cite{Her19bis}, this property is called ``weak liveness''.
  We rename it here, to avoid confusion with our own weak liveness property,
  which is of a very different nature.}  No asset belonging to a compliant party is escrowed forever.
\item \textbf{Strong liveness.}  If all parties are compliant and willing to accept their proposed
  payoffs, then all transfers happen.
\end{itemize}
Here a payoff is \emph{acceptable} to a party $i$ in the deal if party $i$ either receives
all assets $M_{j,i}$ while parting with all assets $M_{i,j}$, or if party $i$ loses nothing at all;
moreover, any outcome where she loses less and/or gains more then an acceptable outcome is also acceptable.

Each entry $M_{i,j}$ contains a type of asset and a magnitude---for instance ``5 bitcoins''.
For each type of asset a separate blockchain is assumed that acts as escrow.
The programming of these blockchains is assumed to be open source, so that all parties can convince
themselves that all escrows abide by the protocol.
With this in mind, their \textbf{Termination} requirement corresponds with ours, 
while \textbf{Safety} is the counterpart of our \textbf{Customer security}.
Our requirement of \textbf{Escrow security} is left implicit in \cite{Her19bis}; since blockchains
do not possess any assets to start with, they surely cannot lose them. Finally, their \textbf{Strong liveness}
property is the counterpart of ours. 

Herlihy, Liskov and Shrira~\cite{Her19bis} offer a timelock commit protocol that requires synchrony,
and assures all three of the above correctness properties. They also offer a certified blockchain
commit protocol that requires partial synchrony and a certified blockchain, and ensures 
\textbf{Safety} and \textbf{Termination};
no protocol can offer \textbf{Strong liveness} in a partially synchronous environment.
For both protocols the correctness is proven for so-called \emph{well-formed} cross-chain deals:
those whose associated directed graph is strongly connected.

In \cite{GGT19} we show that the cross-chain payment cannot be seen as a special kind of cross-chain deal,
nor vice versa.

\paragraph{Acknowledgements}
This research is supported by ARC Discovery Project 180104030: ``Taipan: A Blockchain with
Democratic Consensus and Validated Contracts'' and ARC Future Fellowship 180100496: ``The Red Belly
Blockchain: A Scalable Blockchain for Internet of Things''.

\bibliographystyle{ACM-Reference-Format}
\bibliography{ba-refs}
\end{document}